**Designing the Model Predictive Control for Interval Type-2 Fuzzy T-S Systems Involving Unknown Time-Varying Delay in Both States and Input Vector**

Mohammad Sarbaz[1*]

[1]Electrical and Electronic Engineering Department, Shahed University, Tehran, Iran

[*] Corresponding author. Tel.: +98 9359922712.

E-mail addresses: mohammad.sarbaz@shahed.ac.ir (M. Sarbaz)

*Abstract_* In this paper, the model predictive control is designed for an interval type-2 Takagi-Sugeno (T-S) system with unknown time-varying delay in state and input vectors. The time-varying delay is a weird phenomenon that is appeared in almost all systems. It can make many problems and instability while the system is working. In this paper, the time-varying delay is considered in both states and input vectors and is the sensible difference between the proposed method here and previous algorithms, besides, it is unknown but bounded. To solve the problem, the Razumikhin approach is applied to the proposed method since it includes a Lyapunov function with the original non-augmented state space of system models compared to Krasovskii formula. On the other hand, the Razumikhin method act better and avoids the inherent complexity of the Krasovskii specifically when large delays and disturbances are appeared. To stabilize output results, the model predictive control (MPC) is designed for the system and the considered system in this paper is interval type-2 (IT2) fuzzy T-S that has better estimation of the dynamic model of the system. Here, online optimization problems are solved by the linear matrix inequalities (LMIs) which reduce the burdens of the computation and online computational costs compared to the offline and non-LMI approach. At the end, an example is illustrated for the proposed approach.

## 1. Introduction

One of the most recent subjects that has attracted many attentions in both academic and industry is time-varying delay. This issue has been important for its prominent role in output results and since it has sensible impact on the behavior of systems, many researchers investigate it and its effects when they are designing a process. Therefore, control of systems with time-varying delay in their structures is an essential task and it is necessary to improve output results of the system. the time-varying delay is appeared in different forms and time in the process. It shows itself in different part of the structure and varies the result and even in some cases makes it unstable. Recently, many algorithms and controllers have been proposed to solve this problem with respect to the form of time-varying delay and our needs.

As it was mentioned, since time-varying delay is a significant issue in control theory, many algorithms have been applied to solve this problem. Two well-known approaches that are employed to solve this problem are the Lyapunov–Krasovskii functional (LKF) and the Lyapunov–Razumikhin function (LRF) [1] and [2]. These methods are used for applications of Lyapunov concept for time-delay models. The LKF makes use of an augmentation of the state vector with all delayed states, which makes the applications of Lyapunov approaches to an augmented system with no delay. Hence, the burden of this approach specifically for complex delays, will be unsuitable. On the opposite side, LRF includes a Lyapunov function for the original non-augmented model. However, it is conservative. But suitable for large delays and disturbances. As it is clear, many studies adopted LKF but LRF approaches have not been studied widely [3], [4], and [5]. In [6], an impulsive control is designed for a systems with nonlinear dynamic and distributed time-varying delay. A systems with nonlinear model, time-delay, and disturbances is stabilized by the adaptive control in [7]. In [4], the LRF is employed to encounter the problem of time-varying delay in the system.

With respect to all control algorithms that are designed to stabilize and control output results of system, MPC is a prominent and well-use approach that is employed by many researchers for its convenience and more powerful computing hardware [8] and [9]. The MPC is a controller that its input vector is computed by minimizing a cost function. So, its input vector is considered as an optimal signal vector that used for stabilizing the output. Many times, MPC has been chosen as a alternative to many classic model-based methods and people have used it in their work recently [10], [11], and [12]. In [13], MPC is used for a power electronic system and successfully tracked its output. The MPC is combined with machine learning-based system for a building systems in [14]. The performance of the

AC machine is improved by MPC in [15]. In [16], the prominent MPC is applies to the most used quadcopter system due to its computation and usage and the output of the system is tracked successfully.

Therefore, as it is evident, the model predictive control is applied to many dynamics for its best quality. But an important issue that has turned to a major problem is that all most all of systems have complex dynamic model with high nonlinearity [17]. Hence, it is almost impossible to achieve their dynamic mathematically. A many methods are used to solve this problem like neural networks, reinforcement learning, or fuzzy logic. already, fuzzy systems with **IF-THEN** become well-known and nonlinear and large models are estimated by fuzzy concept [18] and [19]. It can be used for both modeling and as a powerful controller in both academic and industry [20] and [21]. In [22], an adaptive sliding mode controller is applied to a fuzzy model. Robust fuzzy MPC is designed for a fuzzy system in [23]. It can be combined with neural network method and have a better outcome like [24]. In [25], the neural network concept with Kalman filter is synthesized and applied to a fuzzy T-S model. If uncertainties of model are appeared in uncertainties grades of membership functions (MF), approach-based Takagi-Sugeno type- 1 will be conservativeness. So, the IT2 T-S is proposed to solve uncertainties took by the IT2 MF [26] and [27]. The fuzzy MPC is designed for a T-S IT2 large-scale system and besides, uncertainties are considered in [28]. The IT2 fuzzy controller is considered for a power system and shunt active power filter is stabilized correctly in [29].

Obviously, it is trivial that time-varying delay is an uncertain phenomenon and cannot be predicted. It is seen in different part of the structure and its applied time is not specified. It is even can be appeared in the input vector and causes the signal from controller is received by delay. Hence, in this paper we consider a system and controller with the unknown but bounded time-varying delay in both states and input vector. Consequently, the contribution of the paper will be: 1) Designing the fuzzy MPC for system. 2) Considering the IT2 model of the system. 3) Considering the unknown time-varying delay in both states and input vectors. 4) Employing LRF-based approach for discrete-time nonlinear system.

The rest of the paper is: In the section 2 Preliminaries including definitions, lemma, and system description are introduced. Section 3 is for main results, theorems and their proofs. In section 4, the control algorithm is shown. An example of real application to show the effectiveness of the method is proposed in section 5. Finally, the conclusion part is in section 6.

## 2. Preliminaries

### 2.1 Definitions and Lemmas

*Definition 1 [23]:* $R, R_+, Z, Z_+, Z_{[m,n]}, \|x\|, \beta: R_+ \times R_+ \to R_+$, are real, positive real, integers, positive integers, set of integers in the interval $[m,n]$, norm of vector $x_i \in R^c$, M$L$-function ($\beta \in ML$), if for $s > 0, \beta(.,s)$ is a M-function, and for $r > 0, \beta(r,.)$ is decreasing and $\beta(r,s) \to 0$ as $s \to \infty$. And also, A real-valued scalar function $\kappa: R_+ \to R_+$ is an M-function ($\kappa \in M$), if it is continuous, increasing and $\kappa(0) = 0$. So it is true that, $\kappa \in M_\infty$ if $\partial \in M$ and $\lim_{s \to \infty} \kappa(s) = \infty$.

*Lemma:* Due to [30] the following lemma is true:

$$-M_\mu^T Y_\mu^{-1} M_\mu \leq Y_\mu - M_\mu^T - M_\mu \tag{1}$$

### 2.2 System Description

The system is considered:

$$x(k+1) \in f\big(x(k), x_d(k), u(k), u_d(k)\big) \quad k \in Z_+ \tag{2}$$

$x(k), x_d(k) \in X$ and $u(k), u_d(k) \in U$ are the system current states, delayed states, current inputs, and delayed inputs, respectively. And we have:

$$x_d(k) = x(k + d(k)), \quad d(k) \in \mathbb{Z}_{[-h,-1]}. \tag{3}$$
$$u_d(k) = u(k + d(k)), \quad d(k) \in \mathbb{Y}_{[-j,-1]}. \tag{4}$$

where $d(k)$ is the number of delay, $h$ and $j$ are the upper bound of delay, and the minimal delay is 1.

*Definition 2 (Robust Positively Invariant (RPI) set):* For the LRF conditions, consider system (1), a set $\Omega_s$ is an RPI set for the closed-loop system corresponding to the control law, and for the $d(k) \in \mathbb{Z}_{[-h,-1]}$, if $\forall\, x, x_d \in \Omega_s$, and $\forall d(k) \in \mathbb{D}$, the control effort assures that $x^+ \in \Omega_s$.

*Definition 3 (Lyapunov-Razumikhin Function [1]):* Positive definite function $V(x(k))$ is LRF for $x(k+1) = f(x(k), x_d(k))$, if $\mathcal{H}_\infty$-function $\partial_1$, $\partial_2$, and $\mathcal{H}$-function $\rho$ existed in the way that:

$$\partial_1(\| x(k) \|) \leq V(x(k)) \leq \partial_2(\| x(k) \|) \tag{5}$$

$$V(x(k+1)) \leq max\{\bar{V}(x(k)), \rho(\| d(k) \|)\} \tag{6}$$

where $\bar{V}(x(k)) = max\{V(x_i(k)), V(x_d(k))\}$

## 2.3 Time delay T-S system description

The IT2 system is:

$$S^l: \begin{cases} \text{IF } z_1 \text{ is } F_1^l \text{ and } \ldots \text{ and } z_g \text{ is } F_g^l \\ \text{THEN } x(k+1) = A^l x(k) + B^l u(k) + A_d^l x_d(k) + B_d^l u_d(k) \end{cases} \tag{7}$$

in which $A^l, B^l, r_i$ are the system matrices and number of rules. Here, $x(k) \in R^c, u(k) \in R^n$, are state and input vector. The fuzzy system (5) can be shown as:

$$x^+ = A_\mu x + B_\mu u + A_{d\mu} x_d + B_{d\mu} u_d \tag{8}$$

$$A_\mu = \sum_{l=1}^{r_i} \mu^l(z_q) A^l \quad ; \quad A_{d\mu} = \sum_{l=1}^{r_i} \mu^l(z_q) A_d^l$$

$$B_\mu = \sum_{l=1}^{r_i} \mu^l(z_q) B^l \quad ; \quad B_{d\mu} = \sum_{l=1}^{r_i} \mu^l(z_q) B_d^l$$

$\underline{w}^l(z_q(k)) = \prod_{q=1}^{g} \underline{v}_{\tilde{F}_\theta^l}(z_q(k)) \geq 0$ and $\overline{w}^l(z_q(k)) = \prod_{q=1}^{g} \overline{v}_{\tilde{F}_\theta^l}(z_{iq}(k)) \geq 0$ are the lower and upper grades. So, $\underline{v}_{\tilde{F}_\theta^l}(z_q(k)) \in [0,1]$ and $\overline{v}_{\tilde{F}_\theta^l}(z_q(k)) \in [0,1]$ are MF (lower and upper). And, $\overline{v}_{\tilde{F}_\theta^l}(z_q(k)) \geq \underline{v}_{\tilde{F}_\theta^l}(z_q(k))$, hence, $\overline{w}^l(z_q(k)) \geq \underline{w}^l(z_q(k))$.

the following interval sets are the firing strength of rule:

$$W^l = [\overline{w}^l, \underline{w}^l]$$
$$w^l(z_q) = \overline{w}^l(z_q(k)) \bar{\rho}^l(x(k)) + \underline{w}^l(z_q(k)) \underline{\rho}^l(x(k)) \tag{9}$$

here, the lower and upper functions are selected as nonlinear depended to the state variables.
where $\bar{\rho}^l(x(k)) \in [0,1]$ and $\underline{\rho}^l(x(k)) \in [0,1]$ are nonlinear and $\bar{\rho}^l(x(k)) + \underline{\rho}^l(x(k)) = 1$.

*Remark 1:* We can see uncertainties in all nonlinear processes, these uncertainties have uncertain impacts in MF and can consider the lower and upper side of the MF. Sometimes, $\bar{\rho}^l(x(k))$ and $\underline{\rho}^l(x(k))$ are chosen constant. Here, the lower and upper, are selected as nonlinear function`s, depended to state variables. Thus, to facilitate the stability analysis and design of IT2, the lower and upper can be exploited.

The nonlinear fuzzy MPC is:

$$C_i^l: \begin{cases} \text{IF } z_l \text{ is } \tilde{G}_1^l \text{ and } \ldots \text{ and } z_g \text{ is } \tilde{G}_g^l \\ \text{THEN } u(k) = k^l x(k) \end{cases} \tag{10}$$

where $l = 1, 2, \ldots, r_i$, we use the same weight notation $w^l(z_q)$.

$$u(k) = \sum_{l=1}^{r_i} h^l(z_q) k^l x(k) \tag{11}$$

$$u_d(k) = \sum_{l=1}^{r_i} h^l(z_q) k^l x_d(k) \tag{12}$$

in which $\underline{h}^l(z_q(k)) = \prod_{q=1}^g \underline{\sigma}_{\tilde{G}_\theta^l}(z_q(k)) \geq 0$ and $\overline{h}^l(z_q(k)) = \prod_{q=1}^g \overline{\sigma}_{\tilde{G}_\theta^l}(z_q(k)) \geq 0$ are grades of MF. So, $\underline{\sigma}_{\tilde{G}_\theta^l}(z_q(k)) \in [0,1]$ and $\overline{\sigma}_{\tilde{G}_\theta^l}(z_q(k)) \in [0,1]$ are MF (lower and upper). Here, $\overline{\sigma}_{\tilde{G}_\theta^l}(z_q(k)) \geq \underline{\sigma}_{\tilde{G}_\theta^l}(z_q(k))$, therefore, $\overline{h}^l(z_q(k)) \geq \underline{h}^l(z_q(k))$. The following interval sets are the firing strength of rule:

$$h^l(z_q) = \frac{\bar{\mu}^l(z_q(k))\overline{h}^l(x(k)) + \underline{\mu}^l(z_q(k))\underline{h}^l(x(k))}{\sum_{l=1}^{r_i} \bar{\mu}^l(z_q(k))\overline{h}^l(x(k)) + \underline{\mu}^l(z_q(k))\underline{h}^l(x(k))}, \quad H_i^l = [\bar{h}_i^l, \underline{h}_i^l] \tag{13}$$

where $\bar{\mu}^l(x(k)) \in [0,1]$ and $\underline{\mu}^l(x(k)) \in [0,1]$ are nonlinear functions $\bar{\mu}^l(x(k)) + \underline{\mu}^l(x(k)) = 1$.
$\bar{\mu}^l(x(k))$ and $\bar{\mu}^l(x(k))$ relate to state variables instead of constant to lessen conservativeness.

Merge (4) and (8),

$$x(k+1) = \sum_{l=1}^{r_i} \sum_{m=1}^{r_i} w^l(z_q) h^m(z_q) \left( [A^l + B^l k^m] x_i(k) + [A_d^l + B_d^l k^m] x_d(k) \right) \tag{14}$$

### 2.4 Model predictive control

The prediction model is:

$$x(k+t+1|k) = A_\mu x(k+t|k) + A_{d\mu} x_d(k+t|k) + B_\mu u(k+t|k) + B_{d\mu} u(k+t|k) \tag{15}$$

cost function:

$$J(k) = \sum_{n=0}^{T-1} \Pi(k+t|k) + V_t(x(k+T|k)) \tag{16}$$

$\Pi_i(k+t|k)$ and $V_{it}(x_i(k+T|k))$ are stage cost and terminal cost, and $T$ is the prediction duration. $V_{it}(\cdot)$ is positive function. Stage cost:

$$\Pi(k) = x^T(k+t|k)Qx(k+t|k) + u^T(k+t|k)Ru(k+t|k) \tag{17}$$

where $R$ and $Q$ are fixed matrices. Asymptotic convergence to an invariant set is assumed. A positively invariant set is considered called the terminal constraint set that guarantees the system state to enter the terminal set at the end of prediction. The online optimization problem is:

$$\min_{u(k+t|k)} J(k),$$
$$s.t. \quad u(k+t|k) \in U_i$$
$$x(k+T|k) \in \Omega_s$$

where $t \in Z_{[1,T-1]}$ and $\Omega_s$ represent the terminal constraint set. Here, $u \in U := \{u||u_m| \leq u_{m.max}\}$ and $u_m$ is the $m$-th element of the inputs, $m \in Z_{[1,w]}$.

### 3. Main Results

Being the terminal constraint set $\Omega_s$ needs two conditions. First, be a positively invariant set. Later, the positive function $V(x)$ should exist in such wise $\forall\, x \in \Omega_s$,

$$\Omega_s := \{\{x, x_d\} | max\{(x^T P_\mu x), (x_d^T P_\mu x_d)\} \leq \varsigma\} \tag{18}$$

where $P_\mu = \sum_{l=1}^{r_i} w^l(z_q) P$, $\varsigma$ is a positive constant and $\Lambda(x, x_d) = \sum_{l=1}^{r_i} w^l(z_q) k^l x(k)$ is control law.

and the second condition is:

$$\gamma_3(\|x\|) \leq V(x)) \leq \gamma_4(\|x\|) \tag{19}$$

$$V(x^+) - V(x) < -x^T Q x - u^T R u \tag{20}$$

where $\varsigma_3$ and $\varsigma_4$ are $H_\infty$ functions, $V(x)$ is given as:

$$V(x) = \sum_{l=1}^{r_i} w^l(z_q) x^T P x \tag{21}$$

*Theorem 1*: Consider the system (7), if inequalities (22) and (23) are true,

$$\begin{bmatrix} \varrho\lambda & * & * & * & * \\ 0 & \varrho_d\lambda & * & * & * \\ \chi & \chi_d & -Y_l & * & * \\ QM_\mu & 0 & 0 & -\varsigma Q & * \\ RH_\mu & 0 & 0 & 0 & -\varsigma R \end{bmatrix} < 0 \tag{22}$$

$$\begin{bmatrix} Z & * \\ H_\mu^T & M_\mu^T + M_\mu - X \end{bmatrix} \geq 0, \quad Z_{mm} \leq u_{m,max}^2, s \in \mathbb{Z}_{[0,w]} \tag{23}$$

then $\Omega_s$ is a terminal constraint set regarding the terminal cost function $V(x)$. Here, $\lambda = Y_\mu - M_\mu - M_\mu^T$, $\chi = A_\mu M_\mu + B_\mu H_\mu$, $\chi_d = A_{d\mu}M_\mu + B_{d\mu}H_\mu$, $k_\mu = H_\mu M_\mu^{-1}$ and $Z_{mm}$ is the $mm$th diagonal element of a matrix Z. it is evidence that $\varrho + \varrho_d = 1$ and are positive scalers.

*Proof*:

By using the *Lemma 1*:

$$-M_\mu^T Y_\mu^{-1} M_\mu \leq Y_\mu - M_\mu^T - M_\mu$$

and multiply $\{M_\mu^{-T} \quad M_\mu^{-T} \quad 1 \quad 1 \quad 1\}$ and its transpose to both sides of (23):

$$\begin{bmatrix} -\varrho Y_\mu^{-1} & * & * & * & * \\ 0 & -\varrho_d Y_\mu^{-1} & * & * & * \\ \theta & \theta_d & -Y_l & * & * \\ Q & 0 & 0 & -\varsigma Q & * \\ Rk_\mu & 0 & 0 & 0 & -\varsigma R \end{bmatrix} < 0 \tag{24}$$

(24) can be written as:

$$\begin{bmatrix} -\varrho\varsigma Y_\mu^{-1} & * & * & * & * \\ 0 & -\varrho_d\varsigma Y_\mu^{-1} & * & * & * \\ \theta & \theta_d & -\dfrac{Y_l}{\varsigma} & * & * \\ Q & 0 & 0 & -Q & * \\ Rk_\mu & 0 & 0 & 0 & -R \end{bmatrix} < 0 \tag{25}$$

if we consider $\varsigma Y_l^{-1} = P_l$ and $\varsigma Y_\mu^{-1} = P_\mu$:

$$\begin{bmatrix} Q + k_\mu^T R k_\mu - \varrho P_\mu & * & * \\ 0 & -\varrho_d P_\mu & * \\ \theta & \theta_d & -P_l^{-1} \end{bmatrix} < 0 \tag{26}$$

applying the Schur compliment to the inequality we have:

$$\begin{bmatrix} Q + k_\mu^T R k_\mu - \varrho P_\mu & * \\ 0 & -\varrho_d P_\mu \end{bmatrix} + \begin{bmatrix} \theta \\ \theta_d \end{bmatrix}^T P_l [\theta \quad \theta_d] < 0 \tag{27}$$

multiplying $[x^T \quad x_d^T]$ and transpose to both sides:

$$x^T\theta^T P_l\theta x + x^T\theta^T P_l\theta_d x_d + x_d^T\theta_d^T P_l\theta x + x_d^T\theta_d^T P_l\theta_d x_d + x^T Q x + x^T k_\mu^T R k_\mu x - (\varrho x^T P_\mu x + \varrho_d x_d^T P_\mu x_d) < 0 \tag{28}$$

so, we have:

$$(\theta x + \theta_d x_d)^T P_l(\theta x + \theta_d x_d) + x^T Q x + x^T k_\mu^T R k_\mu x - (\varrho x^T P_\mu x + \varrho_d x_d^T P_\mu x_d) < 0 \tag{29}$$

consider $A_\mu + B_\mu k_\mu = \theta$ and $A_{d\mu} + B_{d\mu}k_\mu = \theta_d$

$$\left((A_\mu + B_\mu k_\mu)x + (A_{d\mu} + B_{d\mu}k_\mu)x_d\right)^T P_l \left((A_\mu + B_\mu k_\mu)x + (A_{d\mu} + B_{d\mu}k_\mu)x_d\right) + x^T Q x + x^T k_\mu^T R k_\mu x \\ - (\varrho x^T P_\mu x + \varrho_d x_d^T P_\mu x_d) < 0 \tag{30}$$

$$\left(A_\mu x + B_\mu k_\mu x + A_{d\mu} x_d + B_{d\mu} k_\mu x_d\right)^T P_l \left(A_\mu x + B_\mu k_\mu x + A_{d\mu} x_d + B_{d\mu} k_\mu x_d\right) + x^T Q x + x^T k_\mu^T R k_\mu x$$
$$- \left(\varrho x^T P_\mu x + \varrho_d x_d^T P_\mu x_d\right) < 0 \tag{31}$$

$$x^{+T} P_\mu^+ x^+ + x^T Q x + x^T k_\mu^T R k_\mu x - \left(\varrho x^T P_\mu x + \varrho_d x_d^T P_\mu x_d\right) < 0 \tag{32}$$

if $\varrho + \varrho_d = 1$, so, $\varrho x^T P_\mu x + \varrho_d x_d^T P_\mu x_d \leq V(x)$, and:

$$V(x^+) - V(x) < -x^T Q x - u^T R u \tag{33}$$

the proof of inequality (23) can be seen in [23]. The Proof is completed ∎

## 4. Control Algorithm

To end designing of control, the online control approach illustrated. Later, feasibility and stability will be studied. $\Omega_s$ should satisfy conditions of positively invariant set with respect to (18), illustrates that:

$$V(x) \leq \varsigma \tag{34}$$

since $x, x_d \in \Omega_s$, the nest problem is assumed:

$$\min \varsigma, \quad subject\ to\ \ V(x) \leq \varsigma$$

and the condition for this problem is:

$$\begin{bmatrix} 1 & * \\ x(k+j) & Y_l \end{bmatrix} \geq 0, \quad j \in Z_{[-h,0]}, \quad l \in Z_{[1,L]} \tag{35}$$

**Algorithm**

**Step 1:** Set values of $R, Q, \varrho_d, \varrho$. Obtain the system state $x(k)$.
**Step 2**: Solve the following optimization problem

$$\min_{Y_\mu, Y_l, H_\mu, M_\mu, Z, \varsigma} \varsigma, \tag{36}$$

*subject to* $(22), (23), (35),$„

**Step 2:** Find $k_\mu = H_\mu M_\mu^{-1}$ and apply to the system. Go from $k$ to $k+1$ and then **Step 1**.

*Theorem 2*: For system (7), if the constrained optimization problem has a solution at time 0, then it will always be solvable. Furthermore, recursive feasibility can be achieved.

*Proof:* The proof is in [23]

*Theorem 3:* If the constrained optimization problem is feasible at the initial time 0, then system (7) is stable.

*Proof:* Assume the LRF $V(x(k))$, consider $x(k+\Delta), \Delta \in \{d(k), 0\}$. $\bar{V}(x(k)) = x^T(k+\Delta) P_\mu^* x(k+\Delta)$, in which $P_\mu^* = \sum_{l=1}^{r_i} \mu^l(z_q) P^*$, and $P^*(k)$ is optimal value:

$$\psi_{min}^* \|x(k)\|^2 \leq V(k, x) \leq \psi_{max}^* \|x(k)\|^2 \tag{37}$$

in which
$$\psi_{max}^* = max\{\psi_{max}(P_l^*(k)) | l \in Z_{[1,L]}, k \in R\}$$
$$\psi_{min}^* = min\{\psi_{min}(P_l^*(k)) | l \in Z_{[1,L]}, k \in R\}$$
where $\psi_{max}(\cdot)$ and $\psi_{min}(\cdot)$ are eigenvalues.
and, (20) says that,
$$V_k(x(k+1)) - \bar{V}(x(k)) < -\left(x^T(k) Q x(k) + u^T(k) R u(k)\right) \tag{38}$$
where $V_k(x(k+1)) = x^T(k+1) P_\mu^* x(k+1)$. And if
$$V_k(x(k+1)) - V(k, x) < -x^T(k) Q x(k) \tag{39}$$
(36) at time $(k+1)$, so:

$$V_{k+1}(x(k+1)) \le V_k(x(k+1)) \tag{40}$$

Finally:

$$V_{k+1}(x(k+1)) - V(k,x) < -x^T(k)Qx(k) \tag{41}$$

so the closed-loop system is stable. The proof is completed. ∎

## 5. Example

Now, by a practical application, stirred-tank reactor (CSTR) system, it is proven that the proposed approach is effective. Here, parameters are set as: initial conditions $[0.5 \quad -0.5]$, $Q$ and $R$ $diag\{1e-6 \quad 1e-9\}$ and 0.001, upper bound of delay is 10, $T_s = 0.2$, $-6 \le u \le 6$, $\varrho = 0.8$ and $\varrho_d = 0.2$. the MF is shown by figure 1. The fuzzy system with time-varying delay in both states and input vector is:

*Rule 1:* If $x_2$ is 0.8862 then:

$$A_1 = \begin{bmatrix} 0.75 & 0.0119 \\ -0.2238 & 0.8262 \end{bmatrix}, A_{d1} = \begin{bmatrix} 0.0435 & 0.0003 \\ -0.0061 & 0.0455 \end{bmatrix}, B_1 = \begin{bmatrix} 0.0004 \\ 0.0546 \end{bmatrix}, B_{d1} = 0.001 B_1.$$

*Rule 2:* If $x_2$ is 2.7520 then:

$$A_2 = \begin{bmatrix} 0.6203 & 0.0762 \\ -1.2337 & 1.3265 \end{bmatrix}, A_{d2} = \begin{bmatrix} 0.0403 & 0.0019 \\ -0.0312 & 0.0581 \end{bmatrix}, B_2 = \begin{bmatrix} 0.0023 \\ 0.0698 \end{bmatrix}, B_{d2} = 0.001 B_2.$$

*Rule 3:* If $x_2$ is 4.7052 then:

$$A_3 = \begin{bmatrix} 0.3068 & 0.0442 \\ -3.6621 & 1.0765 \end{bmatrix}, A_{d3} = \begin{bmatrix} 0.0310 & 0.0013 \\ -0.1037 & 0.0528 \end{bmatrix}, B_3 = \begin{bmatrix} 0.0015 \\ 0.0634 \end{bmatrix}, B_{d3} = 0.001 B_3.$$

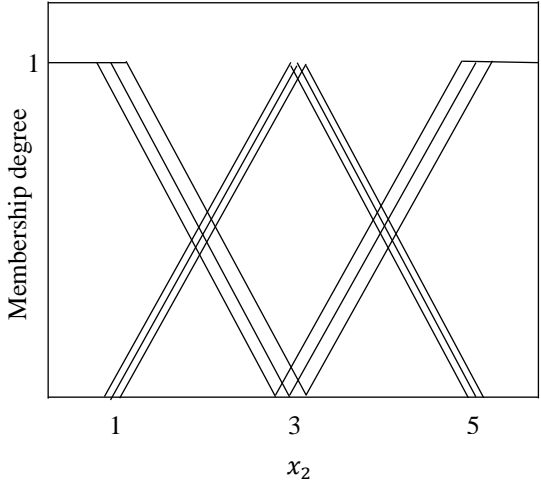

Figure 1. The membership function of IT2

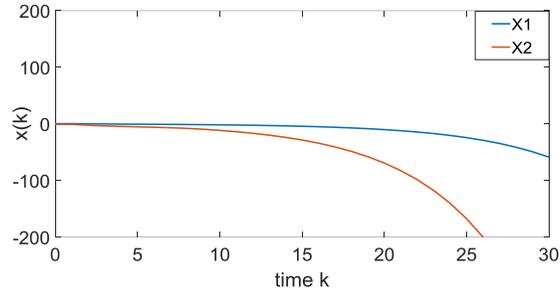

Figure 2. Schematics of states without controller

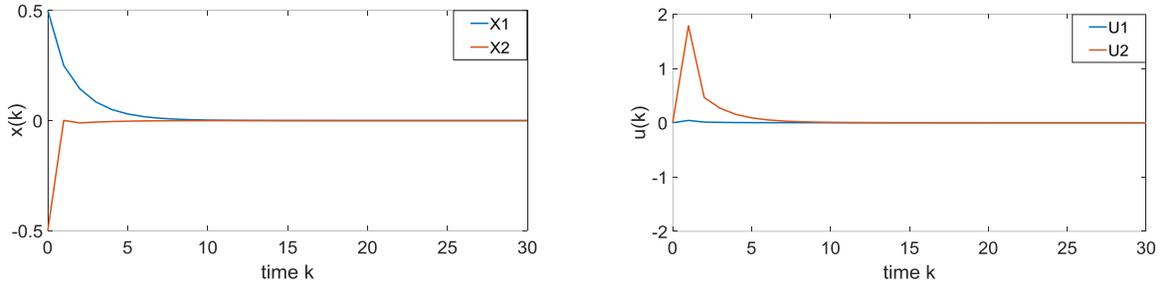

Figure 3. Schematics of states and inputs without delay

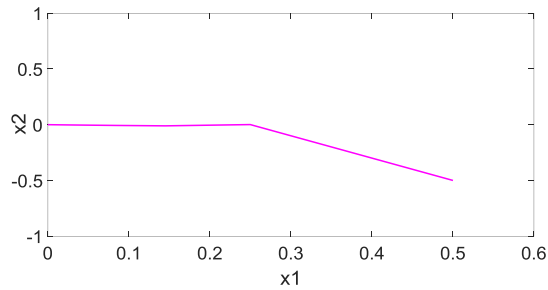

Figure 4. Control performance of the proposed method without delay

*Remark 2:* In this example, three cases are investigated to show the result. Without time-varying delay, with delay in just states, and time-delay in both states and input vector. But before it, in the figure 2 the result of the system and states are illustrated without controller. As it is evident, output results cannot be stabilized and converged to zero without the controller and this shows the effectiveness of the proposed method when it stabilizes the system successfully. Now, different cases are investigated in next remarks.

*Remark 3:* In the first case, the proposed method is applied to the CSTR system without any delay in states and input vector. In this case, as it is clear, states of the system, $[x_1 \quad x_2]$, have a normal and rapid convergence to zero, and stable and perfect results are seen, figure 3. In this figure, although the input restriction is chosen as $-6 \leq u \leq 6$, it can be seen that inputs have a bound of $0 \leq u \leq 1.6$ and after about 5 seconds, when states get inside a bounded region, inputs converge to zero. This means that there is no need to the input vector after some times and shows the optimality of the proposed method. Figure 4 illustrates that system states enter the terminal constraint set finally by applying the proposed approach.

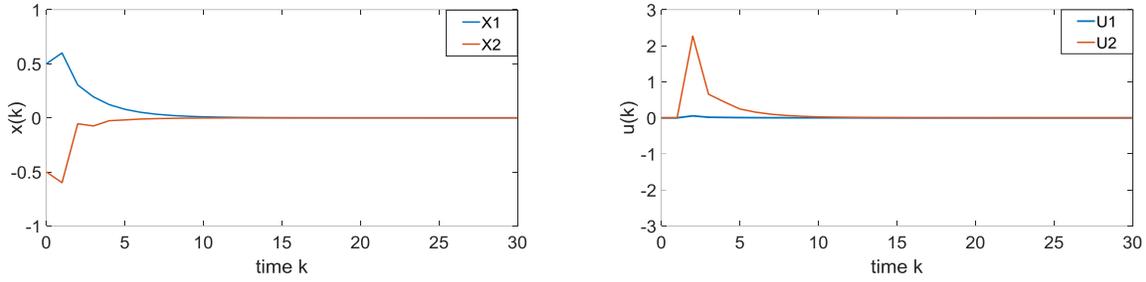

Figure 5. Schematics of states and inputs with delay in states

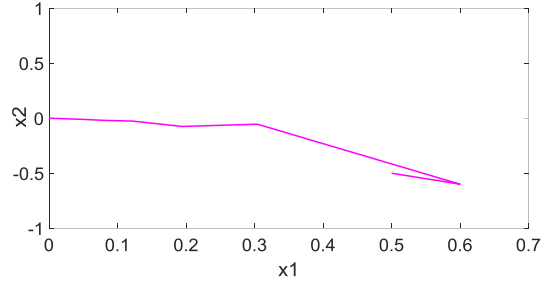

Figure 6. Control performance of the proposed method with delay in states

*Remark 4:* In this case, the time-varying delay is considered just in states and a reverse movement is seen at first step in states, figure 5. By comparing the figure 5 and figure 3, it is trivial that the inputs had more effort to stabilize states. In previous case the bounded was $0 \leq u \leq 1.6$ but in this case it is $0 \leq u \leq 2.3$ and this is for the time-varying delay that is considered in states. Besides, figure 5 shows that states had a reverse movement at first step and this can be seen better in this figure.

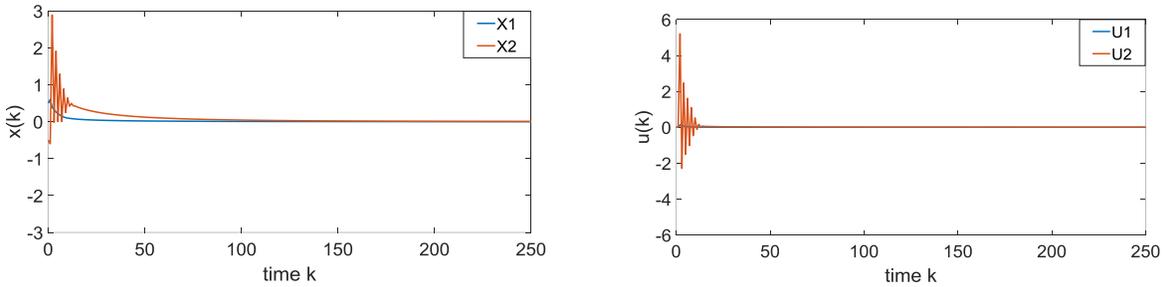

Figure 7. Schematics of states and inputs with delay in both states and inputs

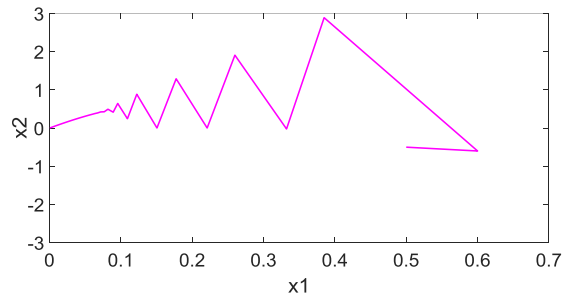

Figure 8. Control performance of the proposed method with delay in both states and inputs

*Remark 5:* In this case, the time-varying delay is considered in both states and input vector. In figure 7, trajectories of both states and input vector are plotted. It is evident that existing the delay in inputs have a sensible effect and the system and as it is clear, the system had some undershoots and overshoots and it took about 70 steps to be converged.

However, it is worthwhile to say that by the proposed approach, finally it was asymptotically converged. Besides, the controller had a hard effort to stabilize states and can be seen in figure 7. The control performance is plotted in figure 8 and shows although it was hard, states were converged in the end. Having to say that existing delay in input vector acts like a strong disturbance but the proposed method is able to encounter it. In this paper, the upper bound of time-varying delay are 10 and this prove the unknown but bounded time-varying delay.

## 6. Conclusion

This manuscript studies design of fuzzy MPC for a IT2 fuzzy system and considerers unknown time-varying delay in both states and input vector. The LRF method is employed for the stabilization of the control due to its simplicity and linearization of the LRF. Gains are computed by minimization of a cost function, and the LMI method is adopted for this. In the MPC approach, the input constraint is considered. By a real example, CSTR, the effectiveness of the method is shown and results are illustrated in three cases, and they are compared. First case is without having delay in system. Second case is for having delay in just states. Third case is for having delay in both states and input vector.